\documentclass
[prl,twocolumn,superscriptaddress,floatfix,showpacs,10pt]{revtex4}%
\usepackage{amsfonts}
\usepackage{amsmath}
\usepackage{amssymb}
\usepackage{graphicx}%
\begin{document}
\title{Yang-Mills as massive Chern-Simons theory: \\ a third way to three-dimensional gauge theories }

\author{Alex S. Arvanitakis}
\email{A.S.Arvanitakis@damtp.cam.ac.uk}
\affiliation{Department of Applied Mathematics and Theoretical Physics, Centre for Mathematical Sciences, University of Cambridge,
Wilberforce Road, Cambridge, CB3 0WA, U.K.}

\author{Alexander Sevrin}
\email{Alexandre.Sevrin@vub.ac.be}
\affiliation{Theoretische Natuurkunde, Vrije Universiteit Brussel, and The International Solvay Institutes, Pleinlaan 2, B-1050 Brussels, Belgium, and 
Physics Department, Universiteit Antwerpen, Campus Groenenborger, 2020 Antwerpen, Belgium}

\author{Paul K. Townsend}
\email{p.k.townsend@damtp.cam.ac.uk} 
\affiliation{Department of Applied Mathematics and Theoretical Physics, Centre for Mathematical Sciences, University of Cambridge,
Wilberforce Road, Cambridge, CB3 0WA, U.K.}

\begin{titlepage}
\begin{flushright}  DAMTP-2015-8 
$\hspace{2.1cm}{}$
\end{flushright}
\vfill

\end{titlepage}

\begin{abstract}

The Yang-Mills (YM) equation  in three spacetime dimensions (3D) can be modified to include a novel parity-preserving  interaction term,  with inverse mass parameter,  in addition to a possible topological mass term. The novelty is that the modified YM equation is  not the Euler-Lagrange equation of any gauge-invariant local action for the YM gauge potential alone.  Instead, consistency is achieved in the ``third way'' exploited by 3D ``minimal massive gravity''.  
We relate our results to the  ``novel Higgs mechanism'' for Chern-Simons gauge theories.

\end{abstract}

\maketitle

\setcounter{equation}{0}

In three spacetime dimensions (3D)  the general gauge-invariant second-order action for a Yang-Mills (YM) gauge potential  $A$ is 
\begin{equation}\label{2ndorder}
I_{TMYM}[A] = \frac{1}{2g^2} \int\!  d^3 x \, \tilde F^\mu\cdot \tilde F_\mu + \frac{\mu}{g^2} \, I_{CS}[A]\, , 
\end{equation}
where $\tilde F$ is the dual Yang-Mills field strength, 
\begin{equation}
\tilde F^\mu = \varepsilon^{\mu\nu\rho} \left(\partial_\nu A_\rho + \frac{1}{2}A_\nu \times A_\rho\right)\, , 
\end{equation}
and $I_{CS}[A] $ is the Chern-Simons action 
\begin{equation}
\frac{1}{2} \int \! d^3 x\, \varepsilon^{\mu\nu\rho}\left[ A_\mu \cdot \partial_\nu A_\rho + \frac{1}{3} A_\mu\cdot A_\nu\times A_\rho\right]\, . 
\end{equation}
Here we suppose, for simplicity, that the gauge group is $SU(2)$ and we use vector algebra notation for products of $SU(2)$ triplets; the generalisation to other gauge groups is straightforward. 

For $\mu=0$,  the action (\ref{2ndorder})  is the  3D YM action with coupling constant $g$. Notice that $g^2$ has dimensions of mass, so that $\mu/g^2$ is dimensionless.   For non-zero $\mu$ the action is that  of ``topologically-massive Yang-Mills'' (TMYM) theory \cite{Schonfeld:1980kb,Deser:1982vy},  which propagates an $SU(2)$ triplet of  spin-1 modes of mass 
 $\mu$. The field equation is
\begin{equation}\label{TMYMeq}
\varepsilon^{\mu\nu\rho} D_\nu \tilde F_\rho + \mu \tilde F^\mu =0\, , 
\end{equation}
where $D_\mu$ is the covariant derivative, defined such that 
\begin{equation}
D_\mu V = \partial_\mu V + A_\mu \times V
\end{equation}
for any $SU(2)$-triplet $V$. 

Let us now add  a source current $J$ to the right hand side of (\ref{TMYMeq}), so that 
\begin{equation}\label{sourced}
\varepsilon^{\mu\nu\rho} D_\nu \tilde F_\rho + \mu \tilde F^\mu = J^\mu \, . 
\end{equation}
Because of the Bianchi identity $D_\mu \tilde F^\mu \equiv 0$, consistency requires the source current to be covariantly conserved:
\begin{equation}
D_\mu J^\mu=0\, . 
\end{equation}
There are two standard ways to construct a source current with this property:
\begin{enumerate}
\item $J=j(\phi)$, the Noether current in a YM background for lower-spin fields $\phi$. In this case $D_\mu j^\mu(\phi)=0$ as a consequence of the $\phi$ equations of motion.

\item $J = \delta I[A]/\delta A$, where $I[A]$ is some gauge-invariant, and Lorentz invariant, functional of $A$. In this case $D_\mu J^\mu \equiv 0$. This will lead to higher-derivative additions to the action. 
\end{enumerate}
There is, however, a third possibility, at least in 3D. In the spin-2 context, this ``third way'' is realised by  ``minimal massive gravity'' (MMG) \cite{Bergshoeff:2014pca,Arvanitakis:2014yja,Arvanitakis:2014xna}, which is a modification of the much-studied ``topologically massive gravity'' \cite{Deser:1981wh}. What we show here is that there is a spin-1 analog of the construction of \cite{Bergshoeff:2014pca}, realised as a particular modification 
of either YM theory (if $\mu=0$) or TMYM theory (if $\mu\ne0$).

Consider the current
\begin{equation}\label{consider}
J^\mu \propto \ \varepsilon^{\mu\nu\rho} \tilde F_\nu \times \tilde F_\rho \, . 
\end{equation}
This current involves only the gauge field $A$, through its field strength. It is {\it not} identically conserved, 
\begin{equation}
D_\mu J^\mu \propto \, \left(\varepsilon^{\mu\nu\rho} D_\mu \tilde F_\nu\right) \times \tilde F_\rho \not\equiv 0\, , 
\end{equation}
but using  the source-free TMYM equation (\ref{TMYMeq}) we find that
\begin{equation}
D_\mu J^\mu\propto \ \mu \tilde F^\mu \times \tilde F_\mu \equiv 0\, . 
\end{equation}
In other words, the third possibility is that $J$ is conserved {\it as a consequence of the YM or TMYM equation itself}.  The obvious difficulty with this idea is that 
we change the YM or TMYM equation as soon as we include $J$ as a source, but in this case 
\begin{equation}
D_\mu J^\mu \propto  J^\mu \times \tilde F_\mu \propto \varepsilon^{\mu\nu\rho} \left(\tilde F_\mu \times \tilde F_\nu\right)\times \tilde F_\rho \equiv 0\, . 
\end{equation}
The final identity is a consequence of the Lie algebra Jacobi identity, so the current $J$ of (\ref{consider}) {\it is} conserved as a consequence of the YM or TMYM equation {\it even after this equation is modified to include $J$}. We have now verified the consistency of the modified  equation 
\begin{equation}\label{MMYM}
\varepsilon^{\mu\nu\rho} \left(D_\nu \tilde F_\rho + \frac{1}{2m} \tilde F_\nu\times \tilde F_\rho\right) + \mu \tilde F^\mu =0\, ,  
\end{equation} 
where $m$ is a further mass parameter.   

It would appear that the new addition to the YM equation breaks parity, even when $\mu=0$, because if the 1-form $A$ is parity even,  as it apparently must be for 
its field strength 2-form $F$ to have definite parity, then the dual 1-form $\tilde F$ is parity odd, implying that its covariant exterior derivative $D\tilde F$ is parity odd but also that  the 2-form $\tilde F\times\tilde F$  is parity even.  Nevertheless, the equation (\ref{MMYM}) does {\it not} break parity when $\mu=0$. To see this  one must  assign  the following 
parity transformation to the $1$-form $A$: 
\begin{equation}
{\cal P}: \ A \to A + m^{-1} \tilde F\, . 
\end{equation}
The parity transformation of $F$ is then 
\begin{equation}
{\cal P}: \ F \to F +\frac{1}{m}\left(D\tilde F + \frac{1}{2m} \tilde F\times \tilde F\right)  \, , 
\end{equation}
so that  $\tilde F$ is still parity odd, and hence  $\tilde F\times \tilde F$ is still parity even, when one uses the $\mu=0$ equation of motion!  
The clash with the apparent odd parity of the $D\tilde F$ term  is resolved by the shift of $A$, which flips the sign of  the $\tilde F\times \tilde F$ term in 
$D\tilde F + \frac{1}{2m} \tilde F\times \tilde F$, so the $\mu=0$ equation of motion preserves parity. We shall see later that  the new YM theory can be formulated in a way that makes this feature manifest. 

The 3D YM stress tensor  is 
\begin{equation}\label{stress}
T_{\mu\nu} = \tilde F_\mu \cdot \tilde F_\nu - \frac{1}{2} \eta_{\mu\nu} \tilde F_\rho \cdot\tilde F^\rho \, . 
\end{equation}
This tensor has the property that $\partial^\mu T_{\mu\nu}=0$ as a consequence of either the YM equation or the TMYM equation. This remains true even if the 
equation used is the modified one of (\ref{MMYM}):
\begin{eqnarray}
\partial^\mu T_{\mu\nu}  &=&  2 \tilde F^\mu \cdot D_{[\mu}\tilde F_{\nu]} = -2m^{-1} \tilde F^\mu \cdot \tilde F_\mu \times \tilde F_\nu \nonumber \\
&=& -2m^{-1} \tilde F^\mu \times \tilde F_\mu \cdot \tilde F_\nu \equiv 0\, . 
\end{eqnarray}
This suggests that the coupling to  3D gravity will be straightforward. 

The same can {\it not}  be said of minimal coupling to lower-spin ``matter''. 
Consider the equation 
\begin{equation}\label{w/source}
\varepsilon^{\mu\nu\rho} \left(D_\nu \tilde F_\rho + \frac{1}{2m} \tilde F_\nu\times \tilde F_\rho\right) + \mu \tilde F^\mu ={\cal J}^\mu\, ,  
\end{equation} 
where ${\cal J}$ is a matter source current.  Taking the divergence of this equation and then using it  to simplify the result, we deduce that 
\begin{equation}\label{cc}
D_\mu {\cal J}^\mu + m^{-1} \tilde F_\mu \times {\cal J}^\mu = 0\, . 
\end{equation}
Only when $m^{-1}=0$ can we take ${\cal J}$ to be a Noether current $j(\phi)$, so it is not immediately clear whether there is a consistent coupling to lower-spin matter. However, given a covariantly conserved matter current $j(\phi)$, and assuming that $m\ne\mu$, the consistency condition (\ref{cc}) is satisfied by a source current of the form
\begin{eqnarray}\label{source}
{\cal J}^\mu &=& j^\mu - \frac{1}{(m-\mu)} \varepsilon^{\mu\nu\rho}D_\nu j_\rho - 
\frac{1}{m(m-\mu)}\varepsilon^{\mu\nu\rho} \tilde F_\nu \times  j_\rho   \nonumber \\    
&& + \quad \frac{1}{2m(m-\mu)^2}\varepsilon^{\mu\nu\rho} j_\nu\times j_\rho\, . 
\end{eqnarray} 
Notice that this is quadratic in the covariantly conserved current $j$, in close analogy to the source tensor for MMG, which is quadratic in the matter stress tensor \cite{Arvanitakis:2014yja}. To verify that (\ref{source}) solves (\ref{cc}) one needs to use the Lie algebra Jacobi identity and  equation (\ref{w/source}), which {\it includes} the source. 

We shall now consider the particular case of coupling to an adjoint Brout-Englert-Higgs (BEH) field; i.e. a triplet scalar $\phi$ for gauge group $SU(2)$.  Assuming that the $\phi$ field equation is 
\begin{equation}
\left[D^\mu D_\mu + 2 V' \right]\phi =0\, ,
\end{equation}
for potential $V(\phi\cdot\phi)$, the covariantly conserved current  is 
\begin{equation}
j_\mu  = \phi \times D_\mu\phi\, . 
\end{equation}
In this case
\begin{eqnarray}
{\cal J}^\mu &=& \phi \times D^\mu \phi - (m-\mu)^{-1}\phi\times (\tilde F^\mu \times \phi)+ \dots  \\
&=& \phi\times \left\{\left[A^\mu- (m-\mu)^{-1} \varepsilon^{\mu\nu\rho}\partial_\nu A_\rho\right] \times \phi\right\} + \dots \nonumber
\end{eqnarray}
where omitted terms are non-linear, even when $\phi$ has a non-zero vacuum value. Let us now suppose that 
\begin{equation}
\phi = v + \varphi\, , 
\end{equation}
where $v$ is a constant  $SU(2)$ triplet and $\varphi$ has zero vacuum value. Then, 
\begin{eqnarray}
v \cdot {\cal J}^\mu &=& 0 + \dots\, , \qquad  \\
v\times {\cal J}^\mu &=& v^2 \left[ v\times A^\mu- (m-\mu)^{-1} \varepsilon^{\mu\nu\rho}\partial_\nu (v \times A_\rho)\right]  + \dots \nonumber 
\end{eqnarray}
where omitted terms are non-linear. We see that the vector potential gauging  the unbroken $U(1)$ gauge group is unaffected by the 
BEH field; it continues to propagate a single spin-1 mode of mass $\mu$.  The other two vector potentials each acquire an explicit 
mass term (with mass-squared $v^2$) so they each propagate a pair of spin-1 modes, but these vector potentials also have
a topological mass term, now with mass parameter $\mu + v^2/(m-\mu)$; notice that this is non-zero even when $\mu=0$. 

In the special case that $\mu=0$ we have, in addition to one massive scalar mode,  one massless mode propagated by the $U(1)$ vector potential and 
four massive modes propagated by the other two vector potentials; each propagates two spin-1 modes of opposite (3D) helicities but with {\it different}
masses because of the topological mass ($v^2/m$) induced by the symmetry breaking. In a parity-preserving theory, massive spin-1 modes must appear in parity-doublets of opposite helicities, so parity is broken by the coupling to matter for finite $m$, even though the source-free theory with $\mu=0$ preserves parity. This is a consequence of the fact that $A$ is not parity inert for finite $m$.  We shall see later how to modify the construction  so
as to preserve parity when $\mu=0$.

Although there is no local gauge invariant action for $A$ alone that yields the source-free equation (\ref{MMYM}), there {\it is} an action involving  auxiliary fields, {\it provided} that $m\ne\mu$.  The Lagrangian density ${\cal L}$ is given  by 
\begin{eqnarray}\label{Lagden}
g^2{\cal L} &=& G_\mu\cdot \tilde F^\mu - \frac{1}{2m}(m-\mu)\,  G_\mu \cdot G^\mu + \mu {\cal L}_{CS}  \\
&+&\!\! \frac{1}{2m}\, \varepsilon^{\mu\nu\rho}\left(G_\mu\cdot D_\nu G_\rho + \frac{1}{3m} G_\mu \cdot G_\nu\times G_\rho\right) .  \nonumber
\end{eqnarray}
In the $m\to\infty$ limit, the auxiliary vector field $G$ (which is also an $SU(2)$ triplet) can be trivially eliminated,  and we are then back to the standard 
TMYM action. More generally, a variation of both $A$ and $G$ induces the following variation of ${\cal L}$:
\begin{eqnarray}\label{variation}
g^2\delta {\cal L} &=& \frac{(m-\mu)}{m}\,  \delta G_\mu \cdot \left(\tilde F^\mu - G^\mu\right) + 
\left(\delta A_\mu + m^{-1}\delta G_\mu\right)\nonumber \\
&&  \cdot \left[\varepsilon^{\mu\nu\rho}\left(D_\nu G_\rho + \frac{1}{2m} G_\nu\times G_\rho\right) + \mu \tilde F^\mu\right] \, . 
\end{eqnarray} 
From this result we see that the field equations imply both  $G^\mu= \tilde F^\mu$ and a further equation that becomes equation (\ref{MMYM}) 
upon substitution for $G$.

We also see from (\ref{variation}) that it is not the $G$ field equation alone that allows us to solve for $G$; that equation also involves $DG$ and a term quadratic in $G$. It is a linear combination of the $A$ and $G$ field equations that allows us to eliminate $G$, but for this reason back-substitution in the {\it action} is illegitimate. This accords nicely with our earlier conclusion  that, as a consequence of its ``third way'' construction, the modified YM equation is not the Euler-Lagrange equation for any local gauge-invariant action constructed from $A$ alone. 

Observe that only the $G_\mu\cdot G^\mu$ term in (\ref{Lagden})  involves the 3D Minkowski metric. From this, and the fact that $G=\tilde F$ on shell, 
it follows that the stress tensor is the usual one, i.e.  $T_{\mu\nu}$ of (\ref{stress}),  times a factor of $(m-\mu)/m$.  We have already verified that 
$\partial^\mu T_{\mu\nu}=0$ remains true for finite $m$. We now see that the energy will be positive or negative according to the sign of $m(m-\mu)$, and that positive energy requires
\begin{equation}
m(m-\mu) >0\, . 
\end{equation}

We may also use  (\ref{Lagden}) to recover  our earlier result (\ref{source}) for the source current ${\cal J}$. 
We just add  to  ${\cal L}$ the interaction term  $-A_\mu\cdot j^\mu$. This is gauge invariant provided that $D_\mu j^\mu=0$, but (as will become clear shortly) it breaks parity when $m$ is finite.  With this term included, the $A$ equation becomes
\begin{equation}
\varepsilon^{\mu\nu\rho}\left(D_\nu G_\rho  + \frac{1}{2m} G_\nu\times G_\rho\right) + \mu \tilde F^\mu= j\, . 
\end{equation}
Recall now that this equation is needed, in addition to the $G$ equation, to determine $G$, so $G$ will acquire a $j$-dependence. In fact, 
\begin{equation}
G^\mu = \tilde  F^\mu + (m-\mu)^{-1} j\, . 
\end{equation}
If this is now substituted into the $A$ equation and all $j$-dependent terms are taken to the right hand side, the result of (\ref{source}) 
is recovered. This construction parallels the construction in \cite{Arvanitakis:2014yja} of the source tensor for MMG.  

We now show how the parity invariance of the action for $\mu=0$ may be made manifest. First we introduce the new gauge potential 
\begin{equation}
\bar A = A + m^{-1}G \, , 
\end{equation}
and then rewrite the action in terms of $A$ and $\bar A$ by using $G= m(\bar A-A)$. The result is 
\begin{eqnarray}\label{CSdiff}
I &=&  \frac{m}{g^2} I_{CS}[\bar A] - \frac{(m-\mu)}{g^2} I_{CS}[A]   \\
&-&\frac{1}{2g^2} m(m-\mu) \int d^3 x\,  \left(\bar A -A\right)_\mu \cdot \left(\bar A -A\right)^\mu\, . \nonumber
\end{eqnarray}
When $m=\mu$ we may  eliminate $A$, trivially, to obtain a Chern-Simons (CS) action for $\bar A$, here for gauge group $SU(2)$, but the
equations of motion of this CS action are {\it not} equivalent to the $m=\mu$ case of  (\ref{MMYM}), for which special case no action is known.

When $\mu=0$ the action (\ref{CSdiff}) preserves parity if  parity is assumed to exchange $A$ with $\bar A$. Then, 
although parity flips the sign of the two CS actions, it also exchanges them,  so their difference is parity even provided their coefficients
sum to zero, which is the case when $\mu=0$.  After elimination of $G$ to recover the new YM field equation of (\ref{MMYM}), the parity transformation 
of $A$ becomes $A\to \bar A = A + m^{-1}\tilde F$.  It should now be clear how we must proceed if we wish minimal coupling to lower-spin matter to preserve parity when $\mu=0$:  we must choose the gauge potential to be the parity-inert  combination
\begin{equation}\label{defC}
C = \frac{1}{2}\left(A + \bar A\right) = A + \frac{1}{2m} G\, ,  
\end{equation}
and then add the parity-preserving interaction term 
\begin{equation}
{\cal L}_{\rm int} = - C_\mu \cdot j^\mu\, , \qquad \partial_\mu j^\mu + C_\mu \times j^\mu=0\, . 
\end{equation}

Since parity flips the sign of $\mu$ in (\ref{CSdiff}), we may assume that $\mu\ge0$. In addition, the {\it field redefinition} $A \leftrightarrow \bar A$  yields the same action but with $m$ replaced by $\mu-m$, so we may further assume $m\ge\mu/2$. Thus $m \ge \mu/2 \ge 0$ may be assumed {\it without loss of generality}, but 
we also need $m>\mu$ for positive energy, in which case (excluding $m=\mu$) we have 
\begin{equation}
m>\mu\ge 0\, , 
\end{equation}
which excludes $m=0$. 

Remarkably, the  $\mu=0$ case of the action (\ref{CSdiff})  has appeared previously \cite{Mukhi:2011jp}, as a model designed to illustrate 
a ``novel Higgs mechanism'' \cite{Mukhi:2008ux}. In this context it arises from a CS theory for gauge group  $SU(n)\times SU(n)$  coupled  to a bi-fundamental Higgs field that breaks $SU(n)\times SU(n)$ to the diagonal $SU(n)$ subgroup, here $SU(2)$.  This construction yields  an additional  singlet massive scalar field, so once this is included the model becomes a CS gauge theory minimally coupled to scalar fields, which is renormalizable as a 3D quantum field theory. 

If  the source-free Lagrangian density (\ref{Lagden}) is rewritten in terms of $C$ rather than $A$, the result for  $\mu=0$ is particularly 
simple:
\begin{equation}\label{AtoC}
g^2{\cal L} = G_\mu \cdot \tilde H^\mu - \frac{1}{2} G_\mu \cdot G^\mu  
+ \frac{1}{24m^2}\, \varepsilon^{\mu\nu\rho} \, G_\mu \cdot G_\nu\times G_\rho\, , 
\end{equation}
where 
\begin{equation}
\tilde H^\mu = \varepsilon^{\mu\nu\rho} \left(\partial_{\mu} C_{\nu} + \frac{1}{2}C_\mu \times C_\nu \right)\, . 
\end{equation}
Parity is manifestly preserved since $C$ is parity even and $G$  is parity odd. This action was also given in \cite{Mukhi:2011jp,Bagger:2012jb}, where it was observed that the field equation for $G$ can be solved recursively, yielding an infinite series expansion in powers of $1/m^2$: 
\begin{equation}
G^\mu = \tilde H^\mu + \frac{1}{8m^2} \varepsilon^{\mu\nu\rho} \tilde H_\nu\times \tilde H_\rho + {\cal O}\left(m^{-4}\right)\, . 
\end{equation}
We may then back-substitute to get an action for $C$ alone. We may also substitute for $G$ in the $C$ equation to get (for $\mu=0$) an equation for $C$ in the form of an  infinite series, but this  series is not explicitly defined and will not converge for all  values of the dual field-strength $\tilde H$. In contrast, our simple equation 
(\ref{MMYM}) for $A$ is, in addition to being more general,  both explicit and defined for all values of $\tilde H$;  we should note here that a special case  ($m=2\mu$) has appeared previously in a related context \cite{Nilsson:2013fya}.

The Lagrangian density (\ref{AtoC}) is a convenient starting point for the construction of the
Hamiltonian formulation for the $\mu=0$ case.  Performing a time-space split we find that 
\begin{eqnarray}
g^2{\cal L} &=& \frac{1}{2}G_0 \cdot G_0 + G_0\cdot \left(B + \frac{1}{8m^2} \varepsilon_{ij} E^i\times E^j\right)\\
&&\!\!\!\!\!\! +\ E^i \cdot \dot C_i + C_0 \cdot \left(\partial_i E^i + C_i\times E^i\right) - \frac{1}{2} E_i \cdot E_i \, ,  \nonumber 
\end{eqnarray}
where a sum over $i=1,2$ is implicit, and
\begin{equation}\label{defB}
E^i = \varepsilon^{ij} G_j \, , \qquad B= \varepsilon^{ij} \left(\partial_i C_j + \frac{1}{2} C_i \times  C_j \right)\, . 
\end{equation}
The auxiliary field $G_0$ may now be trivially eliminated; this  yields
\begin{equation} 
g^2{\cal L} = E^i \cdot \dot C_i + C_0 \cdot D_i E^i - H\, , 
\end{equation}
where the covariant derivative is now defined with gauge potential $C$ and the Hamiltonian is
\begin{equation}
H= \frac{1}{2} E_i \cdot E_i + \frac{1}{2} \left|B +  \frac{1}{8m^2} \varepsilon^{ij} E_i\times E_j\right|^2\, .  
\end{equation}
Here, $| ..|$ is the  $SU(2)$-triplet norm.  We see that the canonical variables $\{C_i,E^i\}$ are subject to the Gauss-law 
constraint $D_i E^i=0$, as in the standard Hamiltonian formulation  of 3D YM theory. The only difference is in the Hamiltonian,
which includes additional terms. Notice, however, that  these are such that  the Hamiltonian remains manifestly positive.

For the generic case of non-zero $\mu$ it is simpler to perform a time-space split in the action (\ref{CSdiff}). 
Provided that $m(m-\mu)\ne0$ we can then trivially eliminate $(A_0-\bar A_0)$ to get 
\begin{eqnarray} 
{\cal L} &=& \frac{(m-\mu)}{2g^2} \varepsilon^{ij} A_i\cdot \dot A_j - \frac{m}{2g^2} \varepsilon^{ij}\bar A_i \cdot\dot{\bar A}_j  \nonumber \\
&& + \frac{1}{g^2} C_0\cdot \left[m\bar B + (m-\mu)B\right] -H\, , 
\end{eqnarray}
where
\begin{eqnarray}
H&=& \frac{1}{2g^2} \left[ m(m-\mu) \left( A_i - \bar A_i\right)\cdot \left(A_i-\bar A_i\right) \right. \nonumber \\
&&\left. +  \ \frac{1}{m(m-\mu)} \ \left|m\bar B - (m-\mu)B\right|^2\right]\, . 
\end{eqnarray}
Here,  $B$ is defined as in (\ref{defB}),  and $\bar B$ is the same but with $\bar A$ instead of $A$. The field $C_0$ is again 
the time component of the parity-even gauge potential $C$, and it is again a Lagrange multiplier for an $SU(2)$-triplet of ``first-class'' 
constraints, which generate $SU(2)$ gauge transformations of the canonical variables. Notice that the Hamiltonian is positive only if  
$m(m-\mu)>0$, as expected from our earlier discussion of the stress tensor. 

We conclude with a comment on the relation of our construction to M-theory. We defer to \cite{Bagger:2012jb} for a review of the relevance to multi M2-brane dynamics  of the action (\ref{CSdiff}) for $\mu=0$. Its relevance for $\mu\ne0$ follows from work of \cite{Gaiotto:2009mv}, where the the sum of the CS levels  was identified with the  Romans mass of  massive IIA supergravity \cite{Romans:1985tz}. In our construction, this sum is proportional to the mass $\mu$ of 
our modified TMYM equation.  This accords with the fact that consistency of the topologically-massive super-D2-brane in a supergravity background implies the field equations of massive IIA supergravity \cite{Bergshoeff:1997cf}.

\noindent\textbf{Acknowledgements}:  We are grateful to Sunil Mukhi for comments on a draft of this paper, for bringing \cite{Nilsson:2013fya} to our attention, and for sending us his unpublished notes, which contain some of the results presented here.  We also thank Stanley Deser and Bengt Nilsson for helpful correspondence. 

A.S.A. and P.K.T. acknowledge support from the UK Science and Technology Facilities Council (grant ST/L000385/1). A.S.A. also acknowledges support from Clare Hall College, Cambridge,  and from the Cambridge Trust.  A.S is supported in part by the Belgian Federal Science Policy Office through the Interuniversity Attraction Pole P7/37, and in part by the ÒFWO-VlaanderenÓ through the project G020714N and by the Vrije Universiteit Brussel through the Strategic Research Program ÒHigh-Energy PhysicsÓ.

\end{document}